\documentclass[a4paper,twocolumn,unpublished,11pt]{quantumarticle}

\pdfoutput=1

\usepackage[T1]{fontenc}
\usepackage[utf8]{inputenc}
\usepackage[english]{babel}
\usepackage{cite}

\usepackage{amsmath,amssymb,amsthm}
\usepackage{dsfont}
\usepackage{physics}
\usepackage[ruled,vlined]{algorithm2e}

\usepackage{graphicx}
\graphicspath{{./images/}}

\usepackage{multirow}

\usepackage{xcolor}
\usepackage{mdframed}

\usepackage{enumitem}
\usepackage{cancel}
\usepackage[normalem]{ulem}
\usepackage{pifont}
\usepackage{orcidlink}
\usepackage{hyperref}

\begin{document}

\title{Local decoder for the toric code via signal exchange}

\author{Louis Paletta \orcidlink{0009-0009-3921-0404}}
\email{louis.paletta@inria.fr}
\affiliation{Laboratoire de Physique de l'Ecole normale supérieure, ENS-PSL, CNRS, Inria, Centre Automatique et Systèmes (CAS), Mines Paris, Université PSL, Sorbonne Université, Université Paris Cité, Paris, France}


\begin{abstract}
    Local decoders provide a promising approach to real-time quantum error-correction by replacing centralized classical decoding, with significant hardware constraints, by a fully distributed architecture based on a simple, local update rule. We propose a new local decoder for Kitaev's toric code: the 2D signal-rule, that interprets odd parity stabilizer measurements as defects, attracted to each other via the exchange of binary signals. We present numerical evidence of exponential suppression of the logical error rate with system size below a threshold, under a phenomenological noise model with data and measurement errors at each iteration. The construction achieves a significantly improved threshold and optimal finite-size scaling relative to hierarchical schemes. It also provides a lightweight alternative to windowed local decoder constructions while maintaining strong performance, thus enabling a streamlined architecture for a two-dimensional local quantum memory.
\end{abstract}

\maketitle

\section{Introduction}
Quantum error-correcting codes protect information from local noise by delocalizing it over a higher-dimensional Hilbert space. Because quantum interactions are limited to be local in space for most physical platform, a natural implementation is based on topological stabilizer codes, where one encodes quantum information by enforcing local constraints on a set of physical qubits placed on a surface.

Encoding information in a quantum error-correcting code, however, is not sufficient to obtain a reliable quantum computation. Among other requirements, one must also be able to detect and correct errors in real time to avoid their accumulation—or, in other words, to suppress the build-up of entropy within the system. In a standard setting, the error-correction mechanism starts by measuring local stabilizers, yielding the error syndrome. The syndrome is then fed to a classical computer running a decoding algorithm, that returns an error consistent with the syndrome: the decoding task is successful if the proposed correction is equal to the initial error up to stabilizer multiplication.
Most current efficient decoders~\cite{edmonds1965paths,dennis2002topological,delfosse2021almost,chamberland2020triangular,kubica2023efficient,berent2024decoding,beni2025tesseractsearchbaseddecoderquantum} are \textit{global decoders}, however, their reliance on centralized classical processors imposes significant hardware constraints as system size increases. This constitutes one of the main obstacles to reaching the scale necessary for fault-tolerant quantum computation.

\subsection{Local decoders}
\begin{figure}
    \centering
    \includegraphics[width=\linewidth]{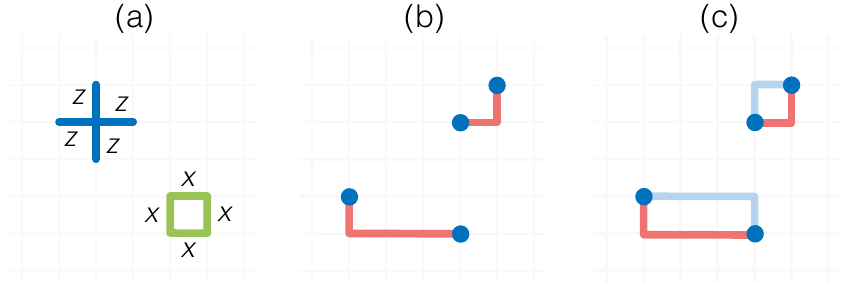}
    \caption[Decoding the toric code]{Decoding the toric code. (a) Physical qubits lie on edges, while $X$-type and $Z$-type stabilizers are associated with plaquettes and stars, respectively. (b) Restricting to bit-flip errors without loss of generality, an initial error—represented as a set of edges—produces a syndrome given by the vertices where parity checks are violated (i.e., defects). (c) Decoding then amounts to pairing these defects, which will be achieved locally by mediating an attractive interaction between them.}
    \label{figure:toric}
\end{figure}

Alternatively, an attractive approach drawing from the theory of cellular automata~\cite{von2017general,moore1956gedanken}, is that of \textit{local decoders}~\cite{kubica2019cellular,vasmer2021cellular,balasubramanian2024local,harrington2004analysis,breuckmann2016local,herold2015cellular,herold2017cellular,michnicki2015towards,lake2025fastofflinedecodinglocal,lake2025localactiveerrorcorrection,paletta2025highperformancelocaldecodersdefect,lang2018strictly,guedes2024quantum,dünnweber2026quantummemoryautonomouscomputation,ueno2021qecool,Chan2026snowflake,Chan2023actisstrictlylocal}, in which each stabilizer measurement site is paired with a small classical processor with limited memory, able to access the value of the site's measurement outcome. A simple local update rule processes the values stored at a site and its neighbors, determines a local Pauli correction, and updates the memory of the site. The decoding task can then be performed by repeatedly applying the local update rule in parallel across all sites. 
The concept is nicely illustrated in the classical setting, where the search for non-ergodic probabilistic cellular automata~\cite{mairesse2014around,lebowitz1990statistical} has led to Toom’s rule~\cite{toom1980stable}: a cellular automaton that preserves a bit of information on a 2D grid for an exponentially long time (scaling with grid size) through local majority votes.
However, because a quantum generalisation is defined on four-dimensional quantum stabilizer codes\cite{kubica2019cellular}, new approach are needed to achieve a functional scheme in low dimensions.
Fortunately, in the quantum setting some constraints can be relaxed: the standard approach assumes noiseless classical processing and permits local memory to scale polylogarithmically with system size for improved performance~\cite{herold2017cellular,harrington2004analysis,breuckmann2016local}. We adopt this framework here.

\subsection{Decoding the toric code}
We focus here on local decoders for the 2D toric code~\cite{kitaev2003fault}, whose variant the surface code remains as of today the leading experimental platforms~\cite{google2025quantum,google2023suppressing,krinner2022realizing}.

In this model, illustrated in Figure~\ref{figure:toric}, the physical qubits lie on the edges of a periodic $d\times d$ lattice, and the $X$ and $Z$ stabilizers act on the four neighboring qubits forming a plaquette and a star, respectively. The code encodes two logical qubits in an $2d^2$-qubit state. Throughout we will assume independent Pauli $X$ and $Z$ errors, so that the problem can be restricted to treating $X$ errors without loss of generality, since $Z$ errors can be treated independently in an analogous manner. The syndrome of an $X$-type error $E$ defined on the edges of the lattice, is represented as a set of vertices $\Sigma = \{\sigma_1, ..., \sigma_p\} \subseteq \mathbb{Z}_{d\times d}$ at which the associated check has odd parity. In the following we will refer to elements of $\Sigma$ as \textit{defects}.

The decoding problem requires matching these defects with a Pauli correction, which is successful when the combined operator of the correction and the error forms a topologically trivial loop on the torus. In contrast to the \textit{offline} regime, which assumes that no errors occur while the decoder is running, stabilizing a quantum memory requires the decoder to operate \textit{online}, where errors may occur between successive applications of the update rules. We focus on the phenomenological error model, where each data qubit independently undergoes a bit flip at each time step with probability $\varepsilon_d$, and each stabilizer measurement outcome is flipped with probability $\varepsilon_m$.
\begin{figure}
    \centering
    \includegraphics[width=0.99\linewidth]{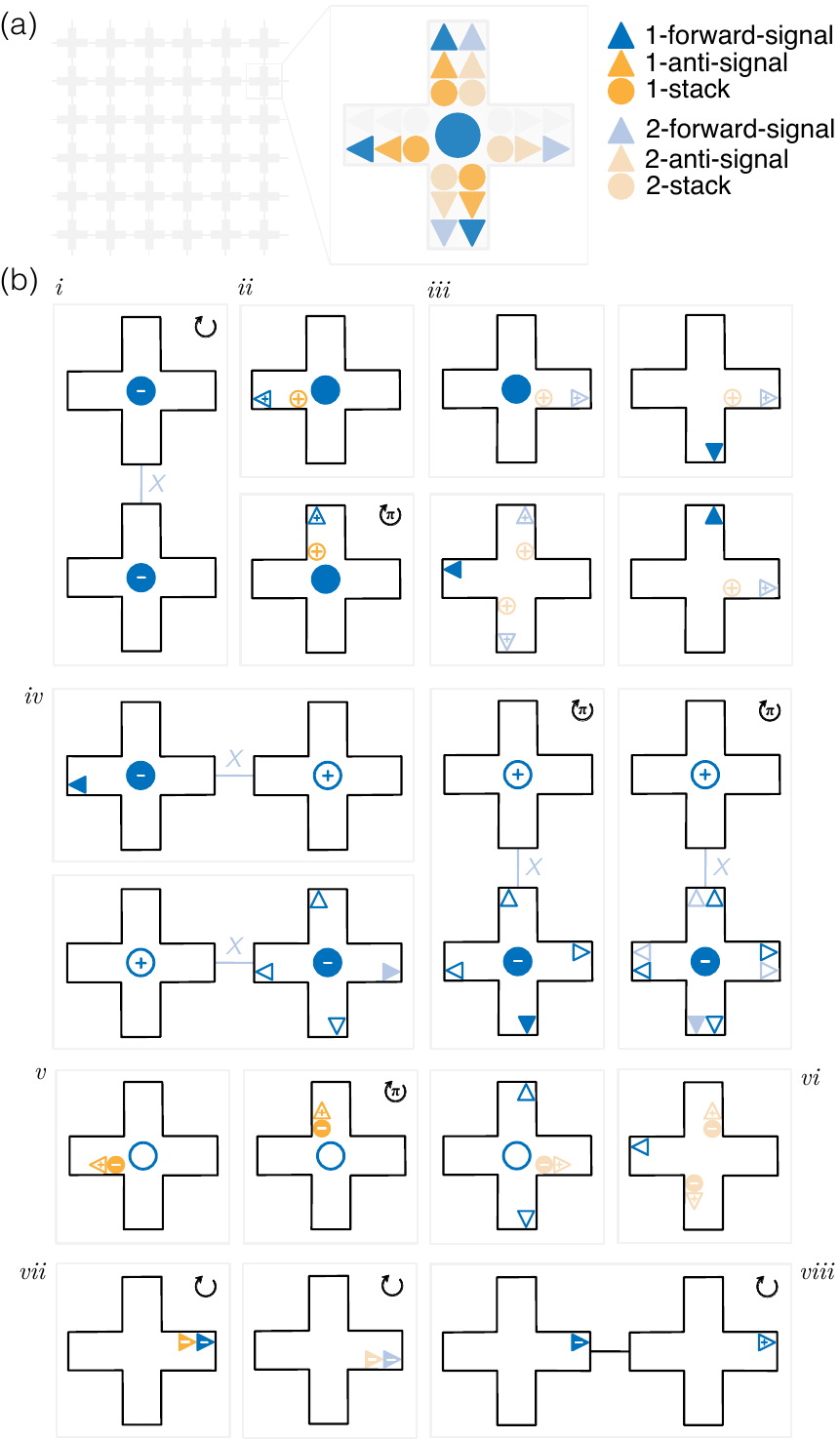}
    \caption{Layout of the 2D signal-rule. (a) Each $Z$ stabilizer site hosts a classical processor storing the stabilizer measurement outcome (i.e., the defect) together with 1- and 2-forward-signals, anti-signals, and stacks for each cardinal direction. Directions unused by the rules (in grey) are retained to define rules by symmetry. (b) Elementary rules of the decoder: filled symbols indicate particles, colored contours indicate empty sites, and plus/minus signs denote updates, subject to implicit constraints from the state space (e.g., no signal creation on an occupied site). A circular arrow marks fourfold rotational symmetry, while $\pi$ denotes symmetry between opposite directions. ($i$) Matching neighboring defects via a Pauli correction (in light blue). ($ii$) Emission of 1-forward-signals from a defect, and associated 1-stack increment. ($iii$) Emission of 2-forward-signals from a 1-forward-signal or a defect, and associated 2-stack increment. ($iv$) Displacement of a defect receiving a forward-signal. ($v$) Emission of 1-anti-signals from the decrement of a non-empty 1-stack in the absence of a defect. ($vi$) Emission of 2-anti-signals from the decrement of a non-empty 2-stack in the absence of a defect and 1-forward-signal. ($vii$) Recombination of a forward-signal with an anti-signal of the same type and direction. ($viii$) Propagation of signals (e.g.\ 1-forward-signals).}
    \label{figure:2d_signal_rules}
\end{figure}
\begin{figure}
    \centering
    \includegraphics[width=0.7\linewidth]{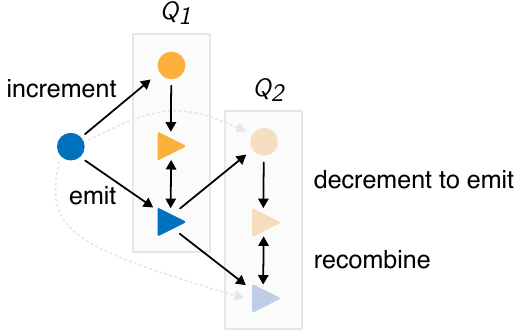}
    \caption{Signal creation diagram, identifying variables of the same type across all cardinal directions. 1-forward-signals are emitted from defects, incrementing the 1-stack, which later decrements to produce 1-anti-signals that recombine with 1-forward-signals. During this process, 1-forward-signals act as defects for 2-forward-signals and 2-stacks in orthogonal directions (certain 2-forward-signals are directly emitted from defects for technical reasons, as indicated in light grey). The pairwise creation and annihilation of signals ensures the conservation of charges $Q_k$, defined by assigning a $+1$ charge to each forward-signal and a $-1$ charge to each stack increment or anti-signal.}
    \label{figure:flow}
\end{figure}
\subsection{Local decoders for the toric code}
Previous proposals for local decoders fall into two main categories (see~\cite{paletta:tel-05488008} for a more detailled review). Some uses a hierarchical structure~\cite{breuckmann2016local,harrington2004analysis,balasubramanian2024local} inspired by classical constructions~\cite{gacs2001reliable,cirel2006reliable}. Others are field-based decoders~\cite{herold2015cellular,herold2017cellular,lake2025fastofflinedecodinglocal,lake2025localactiveerrorcorrection}, where defects are interpreted as particles interacting with each other through a classical field represented by the decoder variables. While some hierarchical constructions can be proven to have a threshold in the online regime~\cite{balasubramanian2024local}, this threshold is in practice relatively low (the best construction achieves $\varepsilon_{c} \simeq 0.13\%$~\cite{breuckmann2016local}). Field-based decoders, in contrast, can achieve strong performance in the offline regime but were not applicable to the online setting~\cite{lake2025fastofflinedecodinglocal} before a windowed structure was introduced in a recent work~\cite{lake2025localactiveerrorcorrection}. 

Alternatively, for the 1D repetition code—to which the 2D toric code is the quantum analogue—an alternative approach has recently been proposed, based on the so-called \textit{signal-rule} decoder, and shown to function in the online regime with good performances~\cite{paletta2025highperformancelocaldecodersdefect}. Similarly to field-based decoders, signal-rule decoders mediate an attractive interaction between defects, but do so through the exchange of several types of point-like particles, referred to as signals, enabling rich dynamics. Its possible generalisation to higher dimensions was left as an open question, which we now answer in the affirmative.

\section{Main results}
\subsection{Overview}
In this work, we introduce a generalisation of signal-rule decoders to the toric code, that we call the \textit{2D signal-rule}. We numerically evaluate its performance in the online regime under a phenomenological noise model with $\varepsilon_d = \varepsilon_m = \varepsilon$, as illustrated in Figure~\ref{figure:logical}. We find that the 2D signal-rule decoder significantly narrows the performance gap with global decoders. Below a pseudothreshold estimated at $\varepsilon_c = 0.68\%$—significantly improved compared to non-windowed scheme—we observe exponential suppression of the logical error rate with increasing system size, achieving near-optimal scaling for practical distances (e.g., $d \lesssim 30$).

While adapting the decoder to incorporate a windowed structure, in the spirit of~\cite{lake2025localactiveerrorcorrection}, could likely improve the threshold at the cost of moderately increased classical resources\footnote{We emphasize that the practical relevance of the exact classical resource requirements depends strongly on the specific implementation, which we do not discuss here.}, the construction presented here has the advantage of operating on a single layer using the exchange of binary signals (rather than integers), thereby requiring minimal storage and a simpler update rule. Focusing only on the width of the update rule, for example in a distance $d=21$ toric code, the number of storage bits needed per site is only $24$, compared to $\sim100$ in~\cite{lake2025localactiveerrorcorrection}. While not optimized for resources usage, these estimates demonstrate that the 2D signal-rule provide a practical lightweight solution for real-time quantum error correction.

\subsection{The 2D signal-rule}
The 2D signal-rule is defined via a transition rule that updates the classical memories associated to each site represented in Figure~\ref{figure:2d_signal_rules}(a), and apply local feedback. In addition to storing the stabilizer measurement outcome as the presence or absence of a defect, for each of the four cardinal directions, four binary variables encode the presence of point-like particles used to propagate information. These particles correspond to \textit{1-} and \textit{2-forward-signals}, and \textit{1-} and \textit{2-anti-signals}. Furthermore, each site maintains \textit{1-} and  \textit{2-stack} registers that are reservoirs of 1- and 2-anti-signals, respectively. A configuration of the decoder at time $t$ is the value of all variables on all sites.

\begin{figure}
    \centering
    \includegraphics[width=\linewidth]{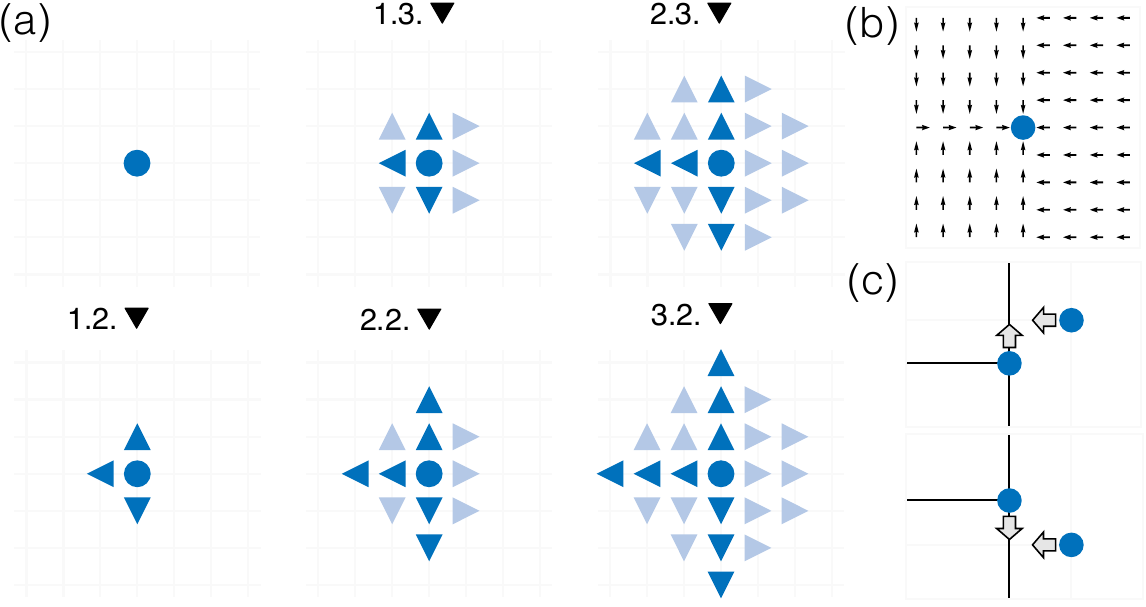}
    \caption{Defect attraction via forward-signal exchange. (a) Attractive interaction from an isolated defect where i.j indicates the step j of iteration i. The attraction is mediated by forward-signals; stacks and anti-signals are omitted for clarity. At each iteration of the decoder, 1-forward-signals (in dark blue) are emitted from the defect and propagate outward, and subsequently 2-forward-signals (in light blue) are emitted from the 1-forward-signals and defects and propagate further outward, generating a 2-dimensional front wave. (b) Attraction basin of a defect, with arrows indicating the direction of attraction for another defect at each site. (c) The asymmetry of the attraction basin ensures agreement between defects displacements during the matching process.}
    \label{figure:field_generation}
\end{figure}
While an offline decoder only needs to correct an initial error, an online decoder must additionally erase the associated syndrome information afterward to free up its limited memory for processing future errors. The 2D signal-rule mediates an effective attractive interaction between defects by generating expanding wavefronts originating from defects through the ballistic propagation of 1- and 2-forward-signals (see Figure~\ref{figure:field_generation}). This induces an attraction between defects, which are eventually matched when they become adjacent. The erasure of the syndrome information from the memory, that is 1- and 2-forward-signals, is implemented using the stack and anti-signal variables. The stack variables count the number of emitted forward-signals (partitionned by directions and types); and eventually decrement to create anti-signals that propagate faster than the forward-signals. These anti-signals recombine with the forward-signals they encounter, effectively canceling them out so that, in the absence of further errors, the decoder eventually relaxes back to the zero configuration.

The relationships between all signals are summarized in Figure~\ref{figure:flow}, from which one can directly observe that, since there is at most one signal of a given type, direction, and nature per site, stack registers only need to store integers up to size $d$ (which can be encoded using $\log d$ bits). Alternatively, we also consider a restricted case where the stack register’s capacity is bounded by some constant integer $m$, and elementary rules are conditionned on not exceeding this bound.

\subsection{Update rule}
An iteration of the 2D signal-rule decoder begins by mapping the local measurement outcomes to the defect variables (step $0.$), followed by the synchroneous application on all site of the elementary rules illustrated in Figure~\ref{figure:2d_signal_rules}(b) and listed below\footnote{We believe the construction to be robust to the asynchroneous case, as long as the conservation relations between signals hold, but do not treat that case here.}:
\begin{enumerate}
    \item \emph{Matching} of neighboring defects by applying a local Pauli correction in between ($i$).

    \item
    \begin{enumerate}[label=\alph*., leftmargin=*, nosep]
        \item \emph{Creation} of 1-forward-signals from defect sites; the associated local 1-stacks are incremented by 1 ($ii$).
        \item \emph{Propagation} of 1-forward-signals by 1 ($viii$).
    \end{enumerate}

    \item
    \begin{enumerate}[label=\alph*., leftmargin=*, nosep]
        \item \emph{Creation} of 2-forward-signals from 1-forward-signals and defect sites; the associated local stacks are incremented by 1 ($iii$).
        \item \emph{Propagation} of 2-forward-signals by 1 ($viii$).
    \end{enumerate}

    \item \emph{Attraction} of defects encountering 1- or 2-forward-signals by applying a local Pauli correction ($iv$).

    \item \begin{enumerate}[label=\alph*., leftmargin=*, nosep]
        \item \emph{Creation} of 1-anti-signals (resp.\ 2-anti-signals) triggered by decrementing the 1-stacks (resp.\ 2-stacks) when no defect is present ($v$) (resp.\ ($vi$)).
        \item (rep. 3 times) \emph{Propagation} by 1 ($viii$) and \emph{recombination} of anti-signals with forward-signals of same type and direction ($vii$).
    \end{enumerate}
\end{enumerate}
\begin{figure*}
    \centering
    \includegraphics[width=\linewidth]{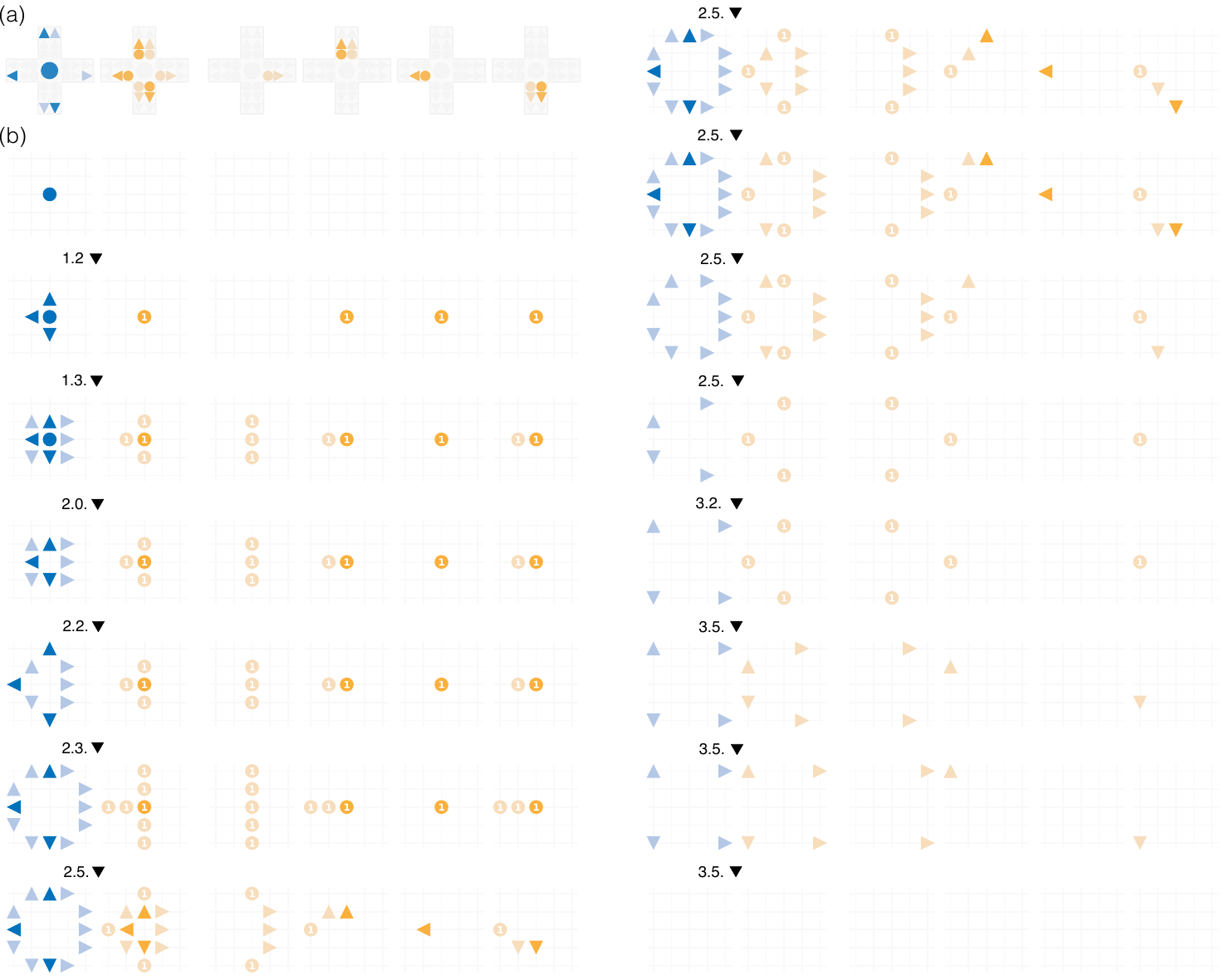}
    \caption{Dynamics following a single initial measurement error. (a) forward-signals and defects are displayed in the first column, while stacks and anti-signals are shown together in the second column and separated by cardinal direction in the four remaining columns. Since only one particle is drawn per site in the second column, particles are displayed according to the following priority order: 1-stacks, 2-stacks, 1-anti-signals, and 2-anti-signals. The maximum stack value over the four cardinal directions is indicated by a white number. (b) Representation of a sequence of configurations of the 2D signal-rule decoder, where steps leaving the configuration unchanged are not shown. In the presence of an initial measurement error, the increment of the stacks upon emission of forward-signals ensure that all signals initially sent eventually recombine at the end.}
    \label{figure:signal2dtime_evolution}
\end{figure*}
For simplicity, this description omits certain auxiliary resources, such as those required to copy signals for local propagation. The full decoding process then consists in repeating the above instructions. 

Animations can be found in~\cite{animation}, and we illustrate the simplest non-trivial case in Figure~\ref{figure:signal2dtime_evolution}, where the forward-signals, emitted from a single defect due to an initial measurement error, are eventually erased along the dynamics. 
In the following we will be interested in the online regime, that is to say in the stability of the memory when errors occur between iterations.

\subsection{Performance}
We numerically evaluate the stability of the 2D signal-rule decoder from Monte-Carlo simulations, under the phenomenological noise model where each data qubits and measurement outcome is flipped independently with probability $\varepsilon$. The automaton is initialized in the zero configuration, and we denote by $P_L(\tau)$ the probability of a logical flip (of any of the two logical qubits) at time $\tau$, determined from a final round of measurement error-free minimum-weight perfect matching decoding~\cite{higgott2022pymatching}. As justified in Section~\ref{section:markov}, $P_L(\tau)$ asymptotically follows a Poisson-like behavior, i.e.\ $P_L(\tau) \propto [1 - (1 - \lambda)^{\tau}]$ for some $\lambda> 0$ from which we define the logical error rate.

\begin{figure}
    \centering
    \includegraphics[width=\linewidth]{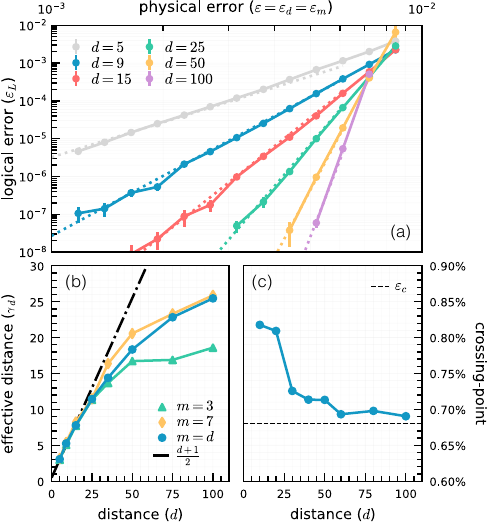}
    \caption{Performance of the $2$D signal-rule decoder.
    (a) Logical error rate as a function of the physical error probability $\varepsilon$ for a phenomenological noise model with $\varepsilon=\varepsilon_d=\varepsilon_m$, shown for several code distances $d$. The logical error rate is obtained by normalizing the failure rate measured in Monte-Carlo simulations.
    The data are fitted using the ansatz $\tfrac{A}{d} (\varepsilon/\varepsilon_c)^{\gamma_d}$, with a distinct exponent $\gamma_d$ for each code distance, which gives $\varepsilon_c = 0.68\%$.
    (b) Effective distance $\gamma_d$ as a function of the code distance $d$, for different stack upper bounds. The optimal scaling in $\tfrac{d+1}{2}$, e.g. reached with minimum-weight perfect matching decoding, is indicated by the black dashed line. (c) For each successive pair of distances $d_i < d_{i+1}$, we fit the modified ansatz 
    $\tilde{A}(\varepsilon / \tilde{\varepsilon}_c)^{\tilde{\gamma}}$ to the corresponding restricted data. This yields a unique crossing point $\tilde{\varepsilon}_c[d_i] = f[d_i, d_{i+1}]$, plotted as a function of $d_i$, and compared to the critical value $\varepsilon_c$ (obtained for all $d$), shown as a dotted line.
    \label{figure:logical}}
\end{figure}
We plot in Figure~\ref{figure:logical}(a) the logical error rate as a function of the physical error rate for different code distances $d$. The decoder performance is quantified by estimating its pseudothreshold and scaling with system size. The pseudothreshold rate is extracted by fitting the logical error rate $\varepsilon_L$ to the ansatz $\tfrac{A}{d} (\varepsilon/\varepsilon_c)^{\gamma_d}$, where the exponent $\gamma_d$ is allowed to depend on the system size. From this fit, we obtain $A = 5.7 \times 10^{-4}$, $\varepsilon_{c} = 0.68\%$, and $\gamma_d$ is plotted as a function of $d$ in Figure~\ref{figure:logical}(b). 

The pseudothreshold significantly improves over hierarchical proposals and diffusive field-based decoders. This is due to the use of ballistically propagating signals, as already demonstrated in~\cite{paletta2025highperformancelocaldecodersdefect,lake2025localactiveerrorcorrection}.
We note that the pseudothreshold does not correspond to a unique crossing point—similarly to many local decoders~\cite{harrington2004analysis,breuckmann2016local,herold2017cellular}. Instead, the crossing point, when estimated from restricting the data to two successive distances, exhibits a drift toward $\varepsilon_c$ for lower error rates as the distance increases, as illustrated in Figure~\ref{figure:logical}(c).

The second performance indicator is the scaling of error suppression with system size, which is characterized by the dependence of $\gamma_d$ on $d$. Since this probes the size of the smallest error leading to a logical failure, we benchmark this effective distance against the optimal $\tfrac{d+1}{2}$ scaling exhibited by many global decoders. We find that $\gamma_d$ closely matches the optimal scaling for system sizes up to $d \lesssim 30$, a regime likely to be most relevant in practice, and for which we show that upper-bounding the stacks to 3 is sufficient to reach comparable performances (amouting to a total classical resource requirement of only 24 bits per site). To go beyond this system size, one must allow the stack to grow with $d$ (interestingly , however, bounded stacks may yield better performance for smaller system sizes). In this regime, the logical error rate continues to decrease exponentially, albeit with a sublinear dependence on the distance. Such suboptimal scaling is characteristic of local decoders, that are typically affected by fractal-like error configurations~\cite{rozendaal2023worst,paletta2025highperformancelocaldecodersdefect,balasubramanian2024local}. Our numerical simulations confirm that this regime persists across all investigated system sizes (up to $d \lesssim 120$).

\section{Markovian dynamics}\label{section:markov}
\subsection{Charge conservation}
While in standard decoding schemes, stabilizer measurement information is erased after a certain time, typically proportional to the code distance. By contrast, in signal-rule decoders, a forward-signal, i.e. a previous odd-parity measurement, does not have a predetermined lifetime. Because of this, it is not clear that the decoder should behave in a Markovian way.

Similarly to the 1D construction however~\cite{paletta2025highperformancelocaldecodersdefect}, the pairwise creation and annihilation of forward-signals and anti-signals or stack increments ensures that the decoder tends to return to the zero configuration after correction of an initial error, thereby allowing for time-invariant dynamics. This property is reminiscent of the presence of conserved quantities arising from pairwise signal creation and recombination, as already illustrated in Figure~\ref{figure:flow}. More precisely, for each signal type $k \in \{1,2\}$, the number of $k$-forward signals minus the sum of $k$-stacks and $k$-anti-signals remains constant. These conserved quantities can furthermore be decomposed into row- and column-wise equivalent for signals associated with the appropriate cardinal directions. We demonstrate numerically in the following that the 2D signal-rule, endowed with this structure, indeed exhibits Markovian dynamics.

\subsection{Numerical evidence}
\begin{figure*}
    \centering
    \includegraphics[width=\linewidth]{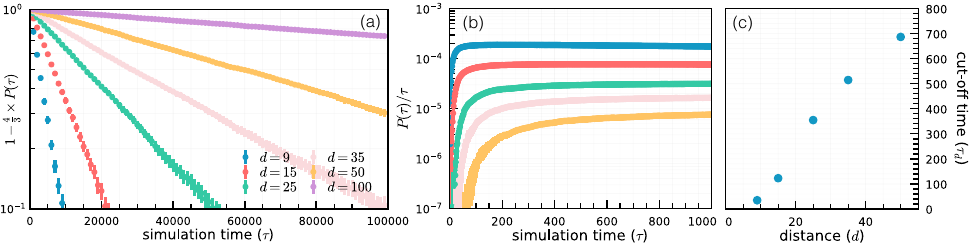}
    \caption{Markovian dynamics.
    (a) $1 - \frac{4}{3} P_L(\tau)$ as a function of the simulation time $\tau$; a constant slope on a logarithmic scale indicates a logical flip probability that is independent of the total simulation time.
    The logical error rate $\varepsilon_L$ is extracted from the asymptotic regime in which $P_L(\tau)/\tau$ reaches a constant value.
    The convergence time to this asymptotic regime is estimated using simulations over shorter time scales, by determining when $P_L(\tau)/\tau$ approaches $\varepsilon_L$. We define the convergence time as $\tau_d := \min \bigl\{ \tau_0 \ge 0 \,\big|\, P_L(\tau)/\tau > 0.9 \cdot \varepsilon_L,\ \forall\, \tau \ge \tau_0 \bigr\}$, which is shown in panel (c). \label{figure:markov}}
\end{figure*}

By defining a logical error rate, one generally assumes that the long-time dynamics of the system can be approximatively captured by a Markov chain over the logical sectors of the code, with the logical error rate $\varepsilon_L$ corresponding to the probability to leave the current state.

Neglecting simultaneous flips of both qubits, the probability to have left the initial state $P_L(\tau)$ by time $\tau$ can be approximated by $P_L(\tau) \simeq \frac{3}{4} [ 1 - \left(1 - \frac{4}{3}\varepsilon_L \right)^\tau ]$, where the factor $3/4$ arises from the four logical sectors of the code. As shown numerically in Figure~\ref{figure:logical}(a), this approximation captures the dynamics of $P_L(\tau)$ well, allowing us to define the logical error rate $\varepsilon_L$. We emphasize that this Poisson-like behavior is enabled by the non-accumulation of signals within the decoder, since any such accumulation could lead to non-Markovian eﬀects.

Finally, for the logical error rate computed from a Monte-Carlo simulations to faithfully represent the system dynamics, the simulation time $\tau$ must be chosen sufficiently large to capture the typical space-time error processes that lead to logical failures. This is illustrated in Figure~\ref{figure:logical}(b), where we examine the convergence of $P_L(\tau)/\tau$ toward $\varepsilon_L$. From this, we estimate the convergence time as $\tau_d \lesssim 10 \times d$, indicating that any longer simulation time suffices to accurately determine $\varepsilon_L$. Accordingly, in all numerical simulations, we set $\tau = 20 \times d$.

\section{Discussion}
In this work, we have introduced a new local decoder for the 2D toric code. Our results demonstrate stability in the online regime and significantly reduce the performance gap with global decoders, making the approach a compelling candidate for real-time quantum error correction in two-dimensional architectures.

It remains an open question whether the observed pseudothreshold represents a genuine threshold. So far, all 2D architectures demonstrating a threshold have relied on hierarchical structures~\cite{balasubramanian2024local,gacs2001reliable,dünnweber2026quantummemoryautonomouscomputation} or logarithmic-depth update rules~\cite{lake2025localactiveerrorcorrection} (note that while the 2D signal-rule requires logarithmic space, its time complexity can likely be made constant, as it consists only of binary operations and constant integer increment/decrement), and it is unclear whether a threshold can be obtained without them. In the present case, a simple argument against the existence of such a threshold comes from considering the matching dynamics of two distant defects. For long-range interaction propagation, it seems that the stacks must avoid decrementing to zero. Otherwise, the forward-signals would be prematurely erased by the emitted anti-signals. This would require a defect to persist at a given site, an event that becomes exponentially unlikely over time, since recurrent noise can move defects, casting doubt on the existence of a genuine threshold\footnote{Interestingly, the 1D version \cite{paletta2025highperformancelocaldecodersdefect} does not appear to suffer from the same instability problem. Long-range attraction remains possible, since signals propagating in a single direction are erased in reverse order of emission.}.
Nevertheless, numerical simulations probing this effect remain inconclusive. It thus remains possible that the two-dimensional signal rule, or one of its variants, could ultimately overcome this limitation. Because a higher dimension generalization of the 2D signal-rule is straightforward, a promising approach consists in a windowed scheme in the spirit of~\cite{lake2025localactiveerrorcorrection}, trading an additional polylog classical overhead for a better threshold, but with an update rule reduced to a constant. We leave this question open.

From a more practical point of view, because of the numerous variant of the 2D signal-rules, we do not claim that our decoders are optimal. However, because the dynamics of the decoder can be naturally understood as interacting particles subject to a global charge conservation, characterization of the underlying structure could provide opportunities for systematic optimization.
Nevertheless, the primary interest in local constructions is their potential to simplify quantum error-correction architecture.
To date, it remains largely unexplored to what extent simple classical computation could be integrated with quantum hardware, with the aim of relaxing connectivity requirements and accelerating quantum error-correction cycles. We hope that these theoretical advances will motivate further investigations. 

\section*{Acknowledgments}
The author thanks Anthony Gandon and Diego Ruiz for numerous valuable comments on the manuscript, Ethan Lake and Nazim Fatès for stimulating exchanges and Mazyar Mirrahimi, Anthony Leverrier and Christophe Vuillot for valuable discussions and collaboration on related works. The author is also grateful to the CLEPS infrastructure from the Inria of Paris for providing resources and support. This work was supported by the Plan France 2030 through the projects NISQ2LSQ (ANR-22-PETQ-0006) and HQI (ANR-22-PNCQ-0002).

\section*{Code availability}
The code used for numerical simulations of the decoder, analysis and visualization is available here~\cite{git}.

\section*{Animation}
Animations in the offline and online regimes can be found here~\cite{animation}.

\bibliographystyle{quantum}
\bibliography{bibliography}

@article{google2025quantum,
  author={Google Quantum AI and Collaborators},
  title={Quantum error correction below the surface code threshold},
  journal={Nature},
  volume={638},
  number={8052},
  pages={920--926},
  year={2025},
  publisher={Nature Publishing Group UK London},
  doi={10.1038/s41586-024-08449-y}
}

@article{google2023suppressing,
  author={Google Quantum AI and Collaborators},
  title={Suppressing quantum errors by scaling a surface code logical qubit},
  journal={Nature},
  volume={614},
  number={7949},
  pages={676--681},
  year={2023},
  publisher={Nature Publishing Group UK London},
  doi={10.1038/s41586-022-05434-1}
}

@article{balasubramanian2024local,
  title   = {A local automaton for the 2D toric code},
  author  = {Balasubramanian, Shankar and Davydova, Margarita and Lake, Ethan},
  journal = {arXiv preprint arXiv:2412.19803},
  year    = {2024},
  doi     = {10.48550/arXiv.2412.19803}
}

@article{breuckmann2016local,
  title   = {Local decoders for the 2D and 4D toric code},
  author  = {Breuckmann, Nikolas P and Duivenvoorden, Kasper and Michels, Dominik and Terhal, Barbara M},
  journal = {arXiv preprint arXiv:1609.00510},
  year    = {2016},
  doi     = {10.48550/arXiv.1609.00510}
}

@article{chamberland2020triangular,
  title     = {Triangular color codes on trivalent graphs with flag qubits},
  author    = {Chamberland, Christopher and Kubica, Aleksander and Yoder, Theodore J and Zhu, Guanyu},
  journal   = {New Journal of Physics},
  volume    = {22},
  number    = {2},
  pages     = {023019},
  year      = {2020},
  publisher = {IOP Publishing},
  doi       = {10.1088/1367-2630/ab68fd}
}

@inproceedings{cirel2006reliable,
  title={Reliable storage of information in a system of unreliable components with local interactions},
  author={Cirel’son, BS},
  booktitle={Locally Interacting Systems and Their Application in Biology: Proceedings of the School-Seminar on Markov Interaction Processes in Biology, Held in Pushchino, Moscow Region, March, 1976},
  pages={15--30},
  year={2006},
  organization={Springer},
  doi={10.1007/BFb0070081}
}

@article{delfosse2021almost,
  title     = {Almost-linear time decoding algorithm for topological codes},
  author    = {Delfosse, Nicolas and Nickerson, Naomi H},
  journal   = {Quantum},
  volume    = {5},
  pages     = {595},
  year      = {2021},
  publisher = {Verein zur F{\"o}rderung des Open Access Publizierens in den Quantenwissenschaften},
  doi       = {10.22331/q-2021-12-02-595}
}

@article{dennis2002topological,
  title     = {Topological quantum memory},
  author    = {Dennis, Eric and Kitaev, Alexei and Landahl, Andrew and Preskill, John},
  journal   = {Journal of Mathematical Physics},
  volume    = {43},
  number    = {9},
  pages     = {4452--4505},
  year      = {2002},
  publisher = {American Institute of Physics},
  doi       = {10.1063/1.1499754}
}

@article{edmonds1965paths,
  title     = {Paths, trees, and flowers},
  author    = {Edmonds, Jack},
  journal   = {Canadian Journal of mathematics},
  volume    = {17},
  pages     = {449--467},
  year      = {1965},
  publisher = {Cambridge University Press},
  nolink    = {}
}

@article{gacs2001reliable,
  title     = {Reliable cellular automata with self-organization},
  author    = {G{\'a}cs, Peter},
  journal   = {Journal of Statistical Physics},
  volume    = {103},
  pages     = {45--267},
  year      = {2001},
  publisher = {Springer},
  nolink    = {}
}

@article{guedes2024quantum,
  title     = {Quantum cellular automata for quantum error correction and density classification},
  author    = {Guedes, Thiago LM and Winter, Don and M{\"u}ller, Markus},
  journal   = {Physical Review Letters},
  volume    = {133},
  number    = {15},
  pages     = {150601},
  year      = {2024},
  publisher = {APS},
  doi       = {10.1103/PhysRevLett.133.150601}
}

@phdthesis{harrington2004analysis,
  title  = {Analysis of quantum error-correcting codes: symplectic lattice codes and toric codes},
  author = {Harrington, James William},
  year   = {2004},
  school = {California Institute of Technology}
}

@article{herold2015cellular,
  title     = {Cellular-automaton decoders for topological quantum memories},
  author    = {Herold, Michael and Campbell, Earl T and Eisert, Jens and Kastoryano, Michael J},
  journal   = {npj Quantum information},
  volume    = {1},
  number    = {1},
  pages     = {1--8},
  year      = {2015},
  publisher = {Nature Publishing Group},
  doi       = {10.1038/npjqi.2015.10}
}

@article{herold2017cellular,
  title     = {Cellular automaton decoders of topological quantum memories in the fault tolerant setting},
  author    = {Herold, Michael and Kastoryano, Michael J and Campbell, Earl T and Eisert, Jens},
  journal   = {New Journal of Physics},
  volume    = {19},
  number    = {6},
  pages     = {063012},
  year      = {2017},
  publisher = {IOP Publishing},
  doi       = {10.1088/1367-2630/aa7099}
}

@article{kitaev2003fault,
  title     = {Fault-tolerant quantum computation by anyons},
  author    = {Kitaev, A Yu},
  journal   = {Annals of Physics},
  volume    = {303},
  number    = {1},
  pages     = {2--30},
  year      = {2003},
  publisher = {Elsevier},
  doi       = {10.1016/S0003-4916(02)00018-0}
}

@article{krinner2022realizing,
  title     = {Realizing repeated quantum error correction in a distance-three surface code},
  author    = {Krinner, Sebastian and Lacroix, Nathan and Remm, Ants and Di Paolo, Agustin and Genois, Elie and Leroux, Catherine and Hellings, Christoph and Lazar, Stefania and Swiadek, Francois and Herrmann, Johannes and others},
  journal   = {Nature},
  volume    = {605},
  number    = {7911},
  pages     = {669--674},
  year      = {2022},
  publisher = {Nature Publishing Group UK London},
  doi       = {10.1038/s41586-022-04566-8}
}

@article{kubica2019cellular,
  title     = {Cellular-automaton decoders with provable thresholds for topological codes},
  author    = {Kubica, Aleksander and Preskill, John},
  journal   = {Physical Review Letters},
  volume    = {123},
  number    = {2},
  pages     = {020501},
  year      = {2019},
  publisher = {APS},
  doi       = {10.1103/PhysRevLett.123.020501}
}

@article{kubica2023efficient,
  title     = {Efficient color code decoders in {$d \ge 2$} dimensions from toric code decoders},
  author    = {Kubica, Aleksander and Delfosse, Nicolas},
  journal   = {Quantum},
  volume    = {7},
  pages     = {929},
  year      = {2023},
  publisher = {Verein zur F{\"o}rderung des Open Access Publizierens in den Quantenwissenschaften},
  doi       = {10.22331/q-2023-02-21-929}
}

@book{michnicki2015towards,
  title     = {Towards self-correcting quantum memories},
  author    = {Michnicki, Kamil},
  year      = {2015},
  publisher = {University of Washington},
  nolink    = {}
}

@inproceedings{rozendaal2023worst,
  title={A worst-case analysis of a renormalisation decoder for Kitaev’s toric code},
  author={Rozendaal, Wouter and Z{\'e}mor, Gilles},
  booktitle={2023 IEEE International Symposium on Information Theory (ISIT)},
  pages={625--629},
  year={2023},
  organization={IEEE},
  doi={10.1109/ISIT54713.2023.10206724}
}

@article{toom1980stable,
  title     = {Stable and attractive trajectories in multicomponent systems},
  author    = {Toom, Andrei L},
  journal   = {Multicomponent random systems},
  volume    = {6},
  pages     = {549--575},
  year      = {1980},
  publisher = {Marcel Dekker New York},
  nolink    = {}
}

@article{vasmer2021cellular,
  title     = {Cellular automaton decoders for topological quantum codes with noisy measurements and beyond},
  author    = {Vasmer, Michael and Browne, Dan E and Kubica, Aleksander},
  journal   = {Scientific reports},
  volume    = {11},
  number    = {1},
  pages     = {2027},
  year      = {2021},
  publisher = {Nature Publishing Group UK London},
  doi       = {doi.org/10.1038/s41598-021-81138-2}
}

@article{berent2024decoding,
  title={Decoding quantum color codes with MaxSAT},
  author={Berent, Lucas and Burgholzer, Lukas and Derks, Peter-Jan HS and Eisert, Jens and Wille, Robert},
  journal={Quantum},
  volume={8},
  pages={1506},
  year={2024},
  publisher={Verein zur F{\"o}rderung des Open Access Publizierens in den Quantenwissenschaften},
  doi={doi.org/10.22331/q-2024-10-23-1506}
}

@misc{git,
  title = {Code is available at \url{https://github.com/lpaletta/local-decoder-2d}},
  year = {2025},
  doi = {doi.org/10.5281/zenodo.17100661}
}

@misc{animation,
  title = {Animations can be found at \url{https://lpaletta.github.io/animation.html}}
}

@article{higgott2022pymatching,
  title={Pymatching: A python package for decoding quantum codes with minimum-weight perfect matching},
  author={Higgott, Oscar},
  journal={ACM Transactions on Quantum Computing},
  volume={3},
  number={3},
  pages={1--16},
  year={2022},
  publisher={ACM New York, NY},
  doi={doi.org/10.1145/350563}
}

@inproceedings{ueno2021qecool,
   title={QECOOL: On-Line Quantum Error Correction with a Superconducting Decoder for Surface Code},
   url={http://dx.doi.org/10.1109/DAC18074.2021.9586326},
   DOI={10.1109/dac18074.2021.9586326},
   booktitle={2021 58th ACM/IEEE Design Automation Conference (DAC)},
   publisher={IEEE},
   author={Ueno, Yosuke and Kondo, Masaaki and Tanaka, Masamitsu and Suzuki, Yasunari and Tabuchi, Yutaka},
   year={2021},
   month=dec, pages={451–456}
   }

@misc{lake2025fastofflinedecodinglocal,
      title={Fast offline decoding with local message-passing automata}, 
      author={Ethan Lake},
      year={2025},
      eprint={2506.03266},
      archivePrefix={arXiv},
      primaryClass={quant-ph},
      url={https://arxiv.org/abs/2506.03266}, 
}

@article{lang2018strictly,
  title={Strictly local one-dimensional topological quantum error correction with symmetry-constrained cellular automata},
  author={Lang, Nicolai and B{\"u}chler, Hans Peter},
  journal={SciPost Physics},
  volume={4},
  number={1},
  pages={007},
  year={2018},
  doi={10.21468/SciPostPhys.4.1.007}
}

@incollection{von2017general,
  title={The general and logical theory of automata},
  author={Von Neumann, John},
  booktitle={Systems research for behavioral science},
  pages={97--107},
  year={2017},
  publisher={Routledge}
}

@article{moore1956gedanken,
  title={Gedanken-experiments on sequential machines},
  author={Moore, Edward F and others},
  journal={Automata studies},
  volume={34},
  pages={129--153},
  year={1956},
  publisher={Princeton},
  url={https://www.torrossa.com/en/resources/an/5573245#page=140}
}

@misc{lake2025localactiveerrorcorrection,
      title={Local active error correction from simulated confinement}, 
      author={Ethan Lake},
      year={2025},
      eprint={2510.08056},
      archivePrefix={arXiv},
      primaryClass={quant-ph},
      url={https://arxiv.org/abs/2510.08056}, 
}

@misc{paletta2025highperformancelocaldecodersdefect,
      title={High-performance local decoders for defect matching in 1D}, 
      author={Louis Paletta and Anthony Leverrier and Mazyar Mirrahimi and Christophe Vuillot},
      year={2025},
      eprint={2505.10162},
      archivePrefix={arXiv},
      primaryClass={quant-ph},
      url={https://arxiv.org/abs/2505.10162}, 
}

@phdthesis{paletta:tel-05488008,
  TITLE = {{Local quantum memories and early fault-tolerant algorithms}},
  AUTHOR = {Paletta, Louis},
  URL = {https://hal.science/tel-05488008},
  SCHOOL = {{PSL University}},
  YEAR = {2025},
  MONTH = Oct,
  TYPE = {Theses},
  HAL_ID = {tel-05488008},
  HAL_VERSION = {v2},
}

@misc{dünnweber2026quantummemoryautonomouscomputation,
      title={Quantum Memory and Autonomous Computation in Two Dimensions}, 
      author={Gesa Dünnweber and Georgios Styliaris and Rahul Trivedi},
      year={2026},
      eprint={2601.20818},
      archivePrefix={arXiv},
      primaryClass={quant-ph},
      url={https://arxiv.org/abs/2601.20818}, 
}

@article{mairesse2014around,
  title={Around probabilistic cellular automata},
  author={Mairesse, Jean and Marcovici, Irene},
  journal={Theoretical Computer Science},
  volume={559},
  pages={42--72},
  year={2014},
  publisher={Elsevier},
  doi={10.1016/j.tcs.2014.09.009}
}

@article{lebowitz1990statistical,
  title={Statistical mechanics of probabilistic cellular automata},
  author={Lebowitz, Joel L and Maes, Christian and Speer, Eugene R},
  journal={Journal of statistical physics},
  volume={59},
  number={1},
  pages={117--170},
  year={1990},
  publisher={Springer},
  doi={10.1007/BF01015566}
}

@misc{beni2025tesseractsearchbaseddecoderquantum,
      title={Tesseract: A Search-Based Decoder for Quantum Error Correction}, 
      author={Laleh Aghababaie Beni and Oscar Higgott and Noah Shutty},
      year={2025},
      eprint={2503.10988},
      archivePrefix={arXiv},
      primaryClass={quant-ph},
      url={https://arxiv.org/abs/2503.10988}, 
}

@article{Chan2026snowflake,
  doi = {10.22331/q-2026-03-20-2033},
  url = {https://doi.org/10.22331/q-2026-03-20-2033},
  title = {Snowflake: {A} {D}istributed {S}treaming {D}ecoder},
  author = {Chan, Tim},
  journal = {{Quantum}},
  issn = {2521-327X},
  publisher = {{Verein zur F{\"{o}}rderung des Open Access Publizierens in den Quantenwissenschaften}},
  volume = {10},
  pages = {2033},
  month = mar,
  year = {2026}
}

@article{Chan2023actisstrictlylocal,
  doi = {10.22331/q-2023-11-14-1183},
  url = {https://doi.org/10.22331/q-2023-11-14-1183},
  title = {Actis: {A} {S}trictly {L}ocal {U}nion–{F}ind {D}ecoder},
  author = {Chan, Tim and Benjamin, Simon C.},
  journal = {{Quantum}},
  issn = {2521-327X},
  publisher = {{Verein zur F{\"{o}}rderung des Open Access Publizierens in den Quantenwissenschaften}},
  volume = {7},
  pages = {1183},
  month = nov,
  year = {2023}
}

\end{document}